\documentclass[a4paper,11pt]{article}
\usepackage[utf8x]{inputenc}
\usepackage{graphicx}
\usepackage{url}
\usepackage{comment}
\usepackage[portuguese,algoruled,longend]{algorithm2e}
\usepackage[top=1.6cm, bottom=1.6cm, left=1.6cm, right=1.6cm]{geometry}
\usepackage{textcomp}
\usepackage{url}
\usepackage{natbib}
\usepackage{color}
\usepackage{subfigure}
\usepackage{graphicx}
\usepackage{natbib}
\usepackage{lipsum}
\usepackage{caption}
\usepackage{setspace}
\usepackage{multirow}
\usepackage{amsmath}
\usepackage{parskip}
\usepackage{tabularx}
\usepackage{amssymb}
\usepackage{fontawesome}
\usepackage{float}
\usepackage{bibentry}

\usepackage[table]{xcolor}
\usepackage{graphicx}
\usepackage{svg}
\usepackage{subfig}
\usepackage{array}
\usepackage{xurl}
\usepackage{pdfpages}

\usepackage{latexsym}
\usepackage{enumitem}

\usepackage{multirow}
\usepackage{amsmath}
\usepackage{tabularx}
\usepackage{float}

\usepackage[table]{xcolor}
\usepackage{graphicx}
\usepackage{svg}
\usepackage{subfig}
\usepackage{array}
\usepackage{enumitem}

\usepackage{booktabs}
\usepackage{mathtools} 
\usepackage{adjustbox} 
\usepackage{stfloats} 
\usepackage{siunitx} 

\begin{document}

\begin{center} \begin{Large} \textbf{The False COVID-19 Narratives That Keep Being Debunked: A Spatiotemporal Analysis}
\end{Large} \end{center} 

\begin{normalsize}
	\begin{center}
\normalsize\textbf{{Iknoor Singh, Kalina Bontcheva, and Carolina Scarton}} \\
	The University of Sheffield, United Kingdom\\
    {\{i.singh, k.bontcheva, c.scarton\}@sheffield.ac.uk}\\
\vspace{0.2cm}

	\end{center}
\end{normalsize}

\section*{Abstract} \vspace*{-0.2cm}

\noindent\textit{The onset of the COVID-19 pandemic led to a global infodemic that has brought unprecedented challenges for citizens, media, and fact-checkers worldwide. To address this challenge, over a hundred fact-checking initiatives worldwide have been monitoring the information space in their countries and publishing regular debunks of viral false COVID-19 narratives. This study examines the database of the CoronaVirusFacts Alliance, which contains 10,381 debunks related to COVID-19 published in multiple languages by different fact-checking organisations. Our spatiotemporal analysis reveals that similar or nearly duplicate false COVID-19 narratives have been spreading in multiple modalities and on various social media platforms in different countries, sometimes as much as several months after the first debunk of that narrative has been published by an International Fact-checking Network (IFCN) fact-checker. We also find that misinformation involving general medical advice has spread across multiple countries and hence has the highest proportion of false COVID-19 narratives that keep being debunked. Furthermore, as manual fact-checking is an onerous task in itself, therefore the need to repeatedly debunk the same narrative in different countries is leading, over time, to a significant waste of fact-checker resources. To this end, we propose the idea of including a multilingual debunk search tool in the fact-checking pipeline, in addition to recommending strongly that social media platforms need to adopt the same technology at scale, so as to make the best use of scarce fact-checker resources.}

\section{Introduction}
\label{saptioIntro}

The COVID-19 pandemic has not only triggered a global health emergency but has also led to the emergence of a worldwide infodemic, commonly referred to as a disinfodemic \citep{Posetti2020}. In 2020, virtually everyone encountered or was exposed to various false claims concerning the origin, transmission, and medical treatments of the coronavirus\footnote{\url{https://www.poynter.org/ifcn-covid-19-misinformation/}}. Numerous studies \citep{limaye2020building,tasnim2020impact} have indicated that a majority of these claims originate on various social media platforms, raising concerns about their authenticity due to the lack of a reliable method for swiftly assessing the credibility of the online content. These unverified claims often fall into the category of misinformation, where the person spreading the claim is unaware of its falsity. Additionally, there is disinformation, involving the intentional spread of false information to deceive \citep{Bontcheva2020}. Both misinformation and disinformation have the potential to inflict significant harm\footnote{\url{https://www.bbc.com/news/world-53755067}}. On the other hand, despite the substantial growth in the number of fact-checking initiatives, these efforts are still unable to effectively mitigate the impact of dis/misinformation in the early stages of its spread due to limited resources \citep{nakov2020can, mcglynn2020misinformation, burel2020co}.

Furthermore, a report \citep{fullfactarticle} by the UK's independent fact-checking organisation FullFact shows that there have been cases where similar narratives disseminated in different countries at different times have been debunked by multiple fact-checking organisations, given that the debunk (or fact-check) for that narrative already existed before. However, the previous study \citep{fullfactarticle} was small-scale and lacked in-depth analysis, a gap we aim to address in this paper. In particular, it is unclear how frequently the same false narratives are spread and debunked across different languages or countries. In this paper, we utilise the International Fact-checking Network (IFCN) CoronaVirusFacts Alliance fact-checks database to find all duplicate debunks of the same false narratives concerning COVID-19. While it is possible that these duplicate debunks were generally published on days that lie in proximity to the publication date of the first debunk, our analysis finds that such duplicates differ by weeks and perhaps even by months from their first appearance. These duplicate debunks usually arise when the same narratives are shared recurrently on various social media platforms in different countries at different times\footnote{\url{https://www.aljazeera.com/news/2020/12/2/trump-releases-video-repeating-debunked-election-fraud-claims}}. Although there could be multiple reasons why people persistently repeat debunked narratives \citep{lewandowsky2012misinformation, ecker2010explicit}, one notable factor is that well-known figures, such as politicians, are known to reiterate false statements consistently \citep{nyhan2010corrections, pillai2021effects}. Another possible reason, which we extensively explore in this paper, is that the debunk published in one language might not be available in another language, preventing the spreader from being aware of its debunk. In particular, we address the following research questions in this paper,

\begin{itemize}
    \item[\textbf{RQ1}] Does the database of COVID-19-related debunks contain duplicate debunks of the same false narrative? In the case of duplicate debunks, what is the temporal gap between them, i.e. can the same false narrative resurface again significantly later and spread unhindered by the platforms’ moderation algorithms in a different language or country?
    \item[\textbf{RQ2}] What are the spatiotemporal characteristics of recurrent debunked narratives, and how do these characteristics differ in terms of country, social media platform, and modality of content?
    \item[\textbf{RQ3}] What types of misinformation is most prevalent and has been debunked by multiple fact-checkers across different countries?
    \item[\textbf{RQ4}] Why integrate a multilingual debunked narrative search tool into the fact-checking pipeline to detect previously debunked narratives in multiple languages?
\end{itemize}
 
In this paper, we uncover numerous cases where similar debunked narratives spread at different times, varying in terms of country, social media platform, and modality of content. These narratives usually stem from an original factually inaccurate claim. Additionally, the recurrent spread of narratives of the same false claim gives rise to debunks from multiple fact-checking organisations in different languages. In this paper, we refer to these as ``duplicate claim debunks'' since they all debunk narratives of the same claim. The term ``debunked narratives'' or ``debunked claims'' refers to false narratives or claims that have undergone prior debunking or have been proven inaccurate by professional fact-checkers. Finally, we identify all such duplicate claim debunks in the IFCN database (Section \ref{saptioMethod}). 

We further investigate the spatiotemporal characteristics of the spread of debunked narratives. The analysis reveals that narratives related to general medical advice are particularly prevalent, having disseminated across multiple countries and been debunked multiple times. For instance, narratives regarding the purported benefits of consuming alkaline-rich food to eliminate coronavirus were initially debunked in Europe. Nevertheless, these narratives persisted, as they were again debunked by fact-checking organisations in Asian, South American, and North American countries. Furthermore, the findings also reveal that Facebook users contribute to most of the misinformation, as the same false narratives keep appearing on the platform, oblivious to the fact that the fact-check articles for those narratives have already been published in the past, either in the same language or in a language different from what the user posts in.

Lastly, there is a growing interest in developing automated fact-checking systems \citep{zhou2020survey, singh2020coherence, thorne2018automated}. In this context, before fact-checking a new claim, it is crucial to prevent the spread of narratives that have already been debunked. For instance, a prior study \citep{reis2020can} on WhatsApp public groups in India and Brazil identified a significant amount of misinformation in the form of images shared within the groups, even after undergoing fact-checking. This recurrent spread of debunked narratives has led to the urgent need for retrieval systems to find fact-checked claims. Recent efforts have been made to address this internal gap \citep{barron2020overview, shaar-etal-2020-known}, with researchers focusing on detecting previously debunked narratives in a monolingual setting. However, this paper underscores the importance of including multilingual debunked narratives in the fact-checking pipeline to determine whether a narrative spreading in one language has already been debunked in the same or a different language (cross-lingual setting). Despite the significance of searching for previously debunked narratives in a multilingual setting, it has largely been overlooked by the research community. Furthermore, given the labour-intensive nature of current fact-checking processes, the ability to search for debunked narratives in a cross-lingual setting can prevent the unnecessary duplication of efforts in debunking the same narratives repeatedly. This approach would allow resources to be allocated more efficiently, enabling the timely fact-checking of other unsubstantiated claims.

In the next section (Section \ref{saptioMethod}), we discuss the method used to perform the analysis. Section \ref{spatifindings} mentions the main finding of this paper and in Section \ref{spatioconclude}, we conclude this paper.


\section{Method}
\label{saptioMethod}

To address the research questions outlined in Section \ref{saptioIntro}, we utilise the CoronaVirusFacts Alliance database led by the IFCN Poynter. The IFCN Poynter database comprises debunks from over 100 organisations in 70 countries, covering around 40 languages. All IFCN fact-checkers adhere to specific principles regarding good practices in debunking. We use the IFCN Poynter\footnote{\url{https://www.poynter.org/ifcn-covid-19-misinformation/}} website to collect all claims that underwent fact-checking in 2020.

We crawl a total of 10,381 claims related to COVID-19 along with their corresponding debunk article page. In addition to the fields provided by the Poynter website\footnote{\url{https://www.poynter.org/wp-content/uploads/2020/05/CORONAVIRUS-FACTS-RFP-Data-Description.pdf}}, we extract the following information fields for each debunked claim on the IFCN Poynter website:

\begin{itemize}
    \setlength{\parskip}{0pt}
    \itemsep0em 
    \item `Claim': Original debunked claim statement from the IFCN Poynter website. 
    \item `Country': List of countries where the claim has spread.
    \item `Fact-checking Organisation': Name of the fact-checking organisation that has debunked the claim. 
    \item `Debunk Link': Link to the fact-checking article about the claim.
    \item `Debunk Language': Language used in the fact-checking article detected using \textit{langdetect} Python library\footnote{\url{https://pypi.org/project/langdetect/}}.
    \item `Debunk Date': Date of publication of the fact-checking article detected using \textit{htmldate} Python library\footnote{\url{https://pypi.org/project/htmldate/}}.
    \item `Social media website': List of websites where claims appeared extracted from fact-checking articles using the JAPE rule \citep{song2021classification}.
    \item `Modality of content': Modality of claims extracted from fact-checking articles using the JAPE rule \citep{song2021classification}.
\end{itemize}

\begin{table}[!htbp]
\tiny
\centering
\resizebox{\columnwidth}{!}{%
\begin{tabular}{p{0.20\linewidth}p{0.06\linewidth}p{0.09\linewidth}p{0.20\linewidth}p{0.06\linewidth}p{0.09\linewidth}}
\toprule
\multicolumn{3}{c}{\textbf{Query Claim Debunk}} & \multicolumn{3}{c}{\textbf{Duplicate Claim Debunk}} \\ \midrule
\multicolumn{1}{c}{\textbf{Claim}} & \multicolumn{1}{c}{\textbf{Debunk Org}} & \multicolumn{1}{c}{\textbf{Date}} & \multicolumn{1}{c}{\textbf{Claim}} & \multicolumn{1}{c}{\textbf{Debunk Org}} & \multicolumn{1}{c}{\textbf{Date}} \\ \midrule
Vitamin C can cure coronavirus. & Détecteur de rumeurs & 2020/04/24 & Vitamin C can cure COVID-19. & JTBC news & 2020/03/04 \\ 
\multicolumn{3}{l}{\multirow{6}{*}{}} & Vitamin C is a miracle cure for the novel coronavirus. & Källkritikbyrån & 2020/03/05 \\ 
\multicolumn{3}{l}{} & Vitamin C prevents coronavirus. & TjekDet.dk & 2020/03/04 \\  
\multicolumn{3}{l}{} & Vitamin C will protect you from the coronavirus. & AFP & 2020/03/13 \\  
\multicolumn{3}{l}{} & Consuming large doses of Vitamin C can stop the spread of coronavirus. & Vishvas News & 2020/03/04 \\  
\multicolumn{3}{l}{} & Vitamin C can “stop” the new coronavirus. & FactCheck.org & 2020/02/12 \\  
\multicolumn{3}{l}{} & The coronavirus can be slowed or stopped with the “immediate widespread use of high doses of vitamin C.” & PolitiFact & 2020/01/27 \\ \midrule
Aborted fetal cells are in the COVID-19 vaccine & Science Feedback & 2020/11/16 & Vaccines, including the one for COVID-19, include aborted fetal tissues. & VoxCheck & 2020/04/28 \\ 
\multicolumn{3}{l}{\multirow{2}{*}{}} & Aborted babies used to develop COVID-19 vaccine & AAP FactCheck & 2020/10/22 \\  
\multicolumn{3}{l}{} & CoronaVac uses cells from aborted fetuses. & Aos Fatos & 2020/07/28 \\ \bottomrule
\end{tabular}
}
\caption{Some examples of query claim debunks and their corresponding duplicate claim debunks. Note 1) Fact-checking organisation of the query claim debunk and duplicate claim debunks is different. 2) Date of publication of the duplicate claim debunk is before the date of publication of the query claim.}
\label{tab:spatioDataSample}
\end{table}

To identify similar debunked narratives, we employ the claim field from the debunks collected earlier to identify semantically similar claims that were debunked by multiple fact-checkers. We formulate this as a retrieval problem, where for each claim field, we conduct a semantic search across all other claims in the dataset. Each claim used as a query is denoted as a ``query claim debunk,'' and their retrieved semantically similar claims are referred to as ``duplicate claim debunks''. 

For retrieval, we initially standardise all references to COVID-19 in the claims (e.g., SARS-CoV-2, COVID-19, 2019-nCoV, COVID) with a unified representation, namely ``coronavirus.'' Following this, we employ a multistage approach \citep{nogueira2019passage, singh2021multistage} involving BM25 Okapi algorithm for initial lexical retrieval and a subsequent neural retrieval stage utilising a state-of-the-art text similarity model based on RoBERTa cross-encoder model\citep{liu2019roberta} to identify semantically similar claims. We ensure robust and reliable data by setting a strict 0.8 similarity score threshold and manually verifying the quality to include only relevant duplicate claim debunks. In addition to this, there are two retrieval constraints: 1) The fact-checking organisation of the query claim debunk is different from the fact-checking organisation of the retrieved duplicate claim debunk 2) The date of publication of the duplicate claim debunk is before the date of publication of the query claim debunk. These constraints ensure that we do not get duplicate cases and only the ones which have the debunks from different fact-checking organisations published in the past. Moreover, the IFCN Poynter\footnote{\url{https://www.poynter.org/wp-content/uploads/2020/05/CORONAVIRUS-FACTS-RFP-Data-Description.pdf}} states that the countries mentioned on the debunked claim webpage are where the falsehood was spreading. Therefore, we infer that the claims which have been debunked at different times are the claims that have been spreading in distinct countries at different times.

Finally, for each query claim debunk, we retrieved $N$ ($\geq$1) duplicate claim debunks. For certain analyses (see Section \ref{spatifindings}), we transformed this from a one-to-many relationship into a one-to-one relation between query claim and duplicate claim debunks.
Table \ref{tab:spatioDataSample} shows examples of query claim debunks and their corresponding duplicate claim debunks.

\section{Findings}
\label{spatifindings}

We divide this section into four parts, where each of the below-mentioned findings addresses the four research questions mentioned in section \ref{saptioIntro} in order.

\subsubsection*{Finding 1. COVID-19 debunks in the IFCN database contain a considerable number of fact-checking articles debunking similar narratives that originate in different countries at different times.}

Out of a total of 10,381 debunks in the IFCN database, we identify 1,070 debunks that already have a debunk about a similar claim from a different fact-checking organisation published in the past. This accounts for 10.3\% of all the debunks in the IFCN database. Throughout this paper, we refer to these 1,070 debunks as ``query claim debunks'' and their duplicate counterparts as ``duplicate claim debunks'' (see Section \ref{saptioMethod}). In other words, for each query claim debunk, we have $N$ ($\geq$1) duplicate claim debunks from different fact-checking organisations published in the past. Please refer to Appendix \ref{gephi} for the cluster plot visualisation for duplicate claim debunks.

\begin{figure}[!htbp]
    \centering
    \includegraphics[width=13cm,height=5.3cm]{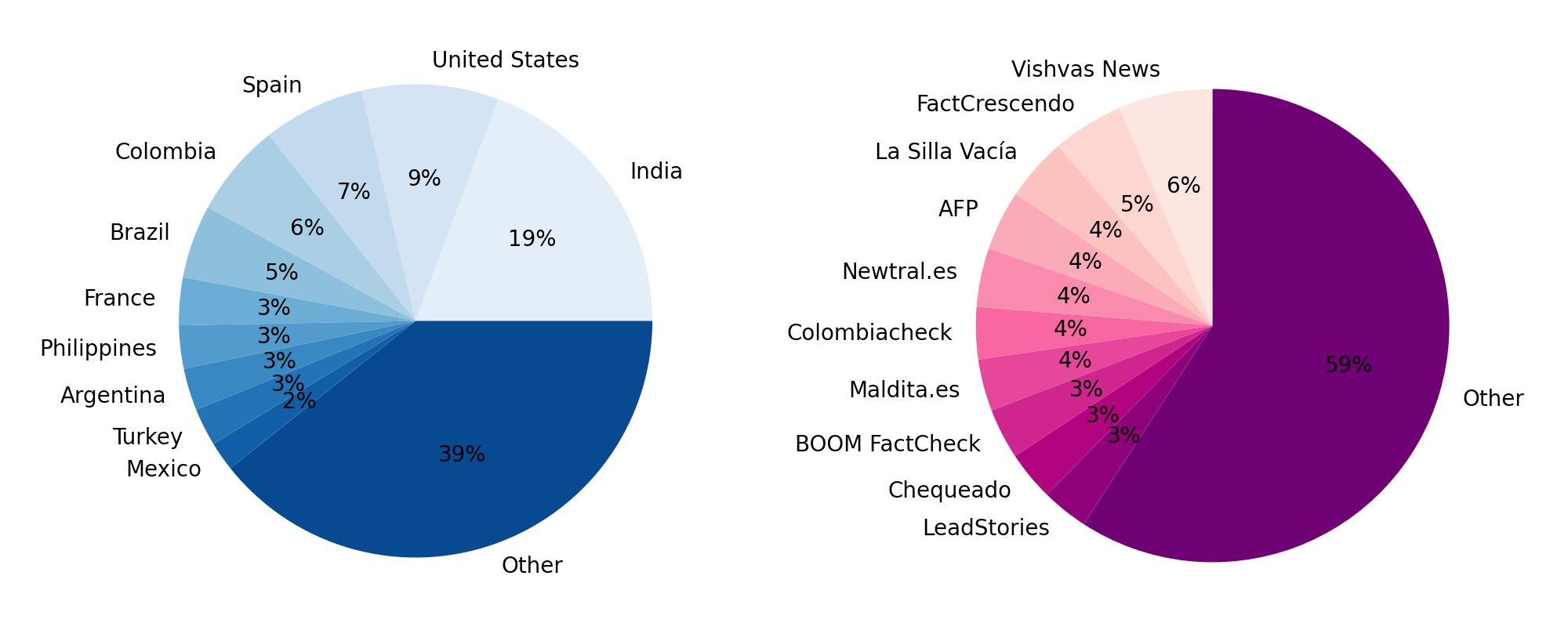}
    \caption{Left: Pie chart distribution for top 10 countries where the claims already debunked were spreading. Right: Pie chart distribution for top 10 fact-checking organisations that published fact-checking articles about the claims that were debunked in the past.}
    \label{fig:spatioFig1}
\end{figure}

Figure \ref{fig:spatioFig1} (left) is the pie chart distribution of the top 10 countries of query claim debunks, i.e., the top countries where claims already debunked are spreading. India and the United States have the largest number of recurring false narratives, and these get debunked multiple times, leading to a waste of fact-checkers' efforts. It indicates that these countries, particularly India with a total proportion of 19\%, are most vulnerable to the spread of narratives that have already been debunked in the past. In general, this also suggests a lack of awareness among the people about prior fact-checked information. 

Figure \ref{fig:spatioFig1} (right) illustrates the pie chart distribution of the top 10 fact-checking organisations of query claim debunks, i.e. the top fact-checking organisations that are debunking narratives for which debunks already existed in the past. The results align with Figure \ref{fig:spatioFig1} (left), where Vishvas News, an Indian fact-checking website, publishes a large number of debunks about previously fact-checked claims.

\begin{figure}[!htbp]
    \centering
    \includegraphics[width=8cm,height=5cm]{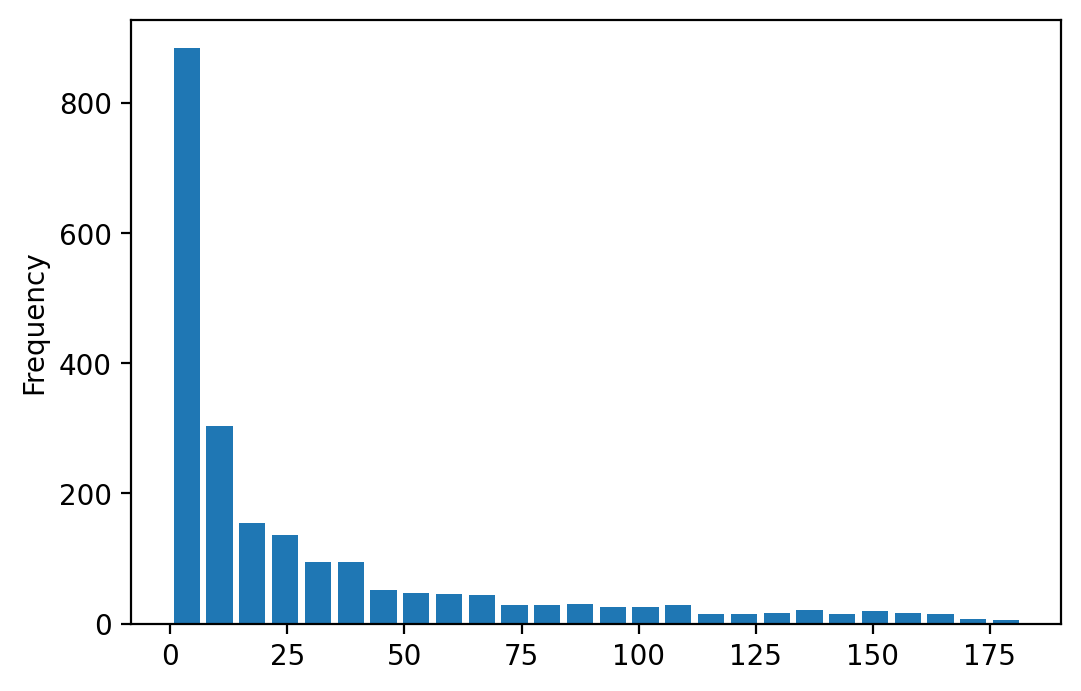}
    \caption{Histogram plot for days difference between query claim debunks and duplicate claim debunks. (Bin set at an interval of 1 week).}
    \label{fig:spatioFig2}
\end{figure}

The difference in days between the publication date of query claim debunks and the duplicate claim debunks is depicted in Figure \ref{fig:spatioFig2}. The histogram plot shows the weekly count with the bin interval set at 7 days. For instance, the first bar indicates that there are 884 cases where the publication date difference between query claim debunk and duplicate claim debunk is one week or less. Similarly, the second bar shows nearly 300 cases with a fortnight difference, and so forth. This reveals that misinformation persists and gets debunked multiple times even after relevant debunks are already available. This is worrisome and the subsequent findings help us understand the reasons for the existence of such duplicate claim debunks.

\subsubsection*{Finding 2. Spatiotemporal characteristics of similar false narratives and their transition between countries, social media platforms and modalities of content.}

\begin{figure}[!htbp]
    \centering
    \includegraphics[width=\textwidth,height=4cm]{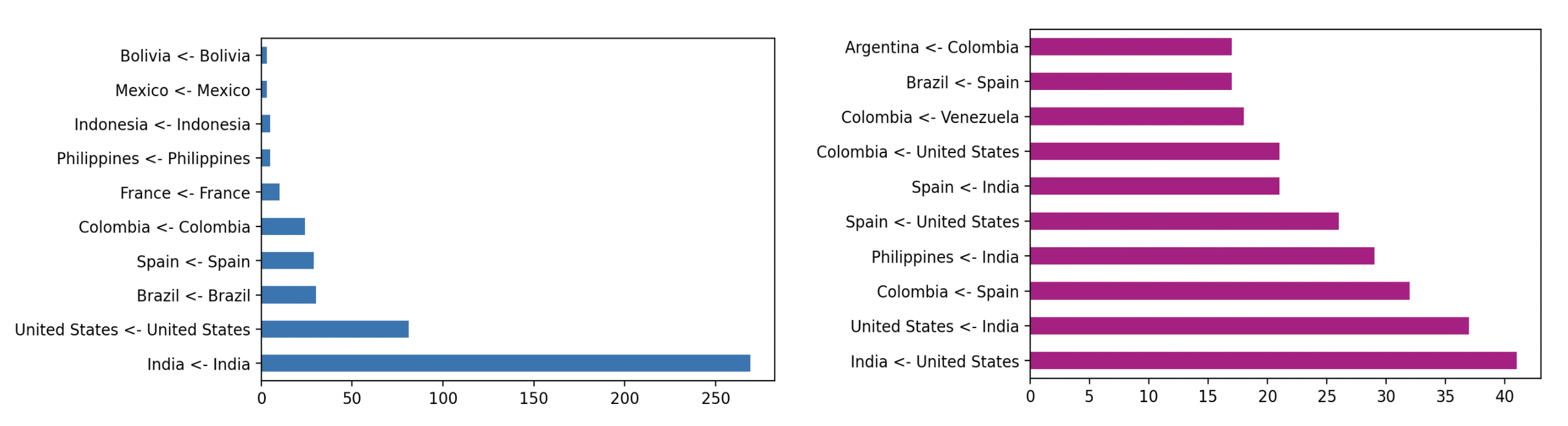}
    \caption{The movement of similar false claims between different country pairs. The bar chart on the left shows the top 10 counts of cases where both the countries are same and the bar chart on the right depicts the top 10 cases where both countries are different.}
    \label{fig:spatioFig3}
\end{figure}

The spatiotemporal characteristics of both query claim debunks and duplicate claim debunks can help reveal how information flows or changes between different debunks. In Figure \ref{fig:spatioFig3}, pie charts illustrate the movement of similar false claims between different countries. For simplicity, we only consider the top 10 country pairs, where Figure \ref{fig:spatioFig3} (left) shows the count of cases where both countries are the same, and Figure \ref{fig:spatioFig3} (right) shows cases where both countries are different.

Since the date of publication of the duplicate claim debunk is before the publication date of the query claim debunk (see Section \ref{saptioMethod}), the symbol ``$\leftarrow$'' between the countries can be treated as the flow of false claims between different country pairs. For example, ``India $\leftarrow$ United States'' indicates that there are around 40 cases where the flow of false claims is from the United States to India. We find that the movement of similar false claims is highest between India and the United States, followed by movement from Spain to Columbia. The conceivable reason for this could be the common language of English and Spanish, respectively, for each of the cases.

\begin{figure}[!htbp]
    \centering
    \includegraphics[width=14cm,height=4cm]{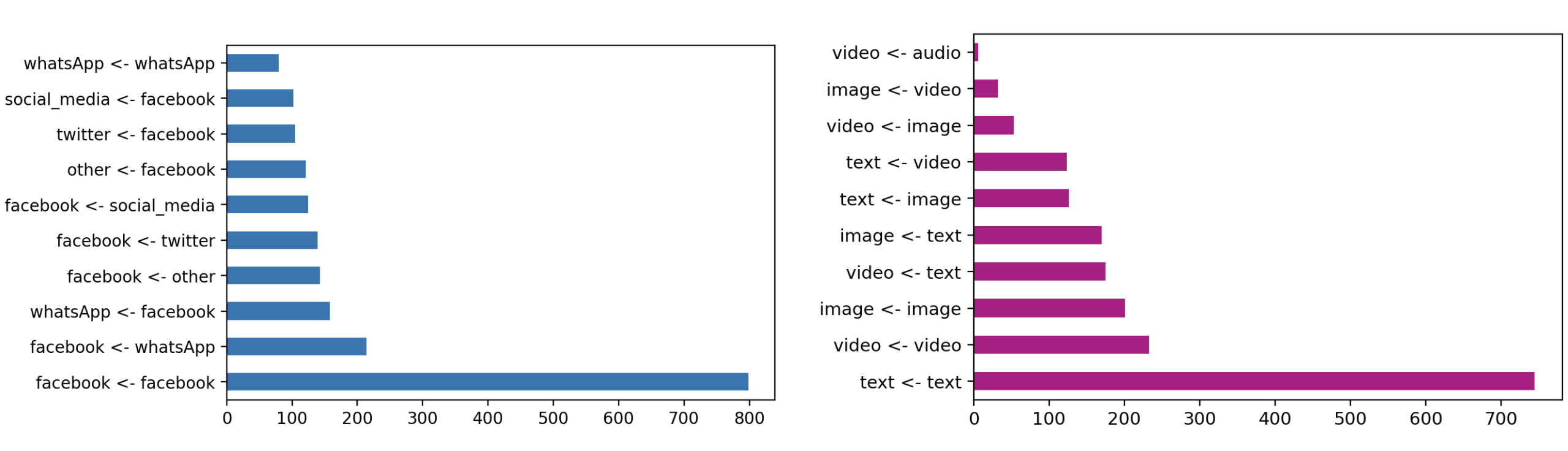}
    \caption{Left: Transition in social media platforms. Right: Transition between modality of content.}
    \label{fig:spatioFig4}
\end{figure}

Figure \ref{fig:spatioFig4} (left) illustrates the change in social media platforms of the claims fact-checked in both query claim debunk and duplicate claim debunk. In other words, it provides insights into the movement of similar false claims from one social media website to another. It suggests that for similar claims, the spread within Facebook itself is the highest, with around 800 cases, followed by occurrences from WhatsApp to Facebook, which has just over 200 instances. This is particularly concerning given that Facebook, increasingly used as a primary source of news \citep{bridgman2020causes}, allows the wide dissemination of content whose falsity has already been fact-checked in the past. 

According to a Pew Research report of 2020\footnote{\url{https://www.pewresearch.org/journalism/2021/01/12/news-use-across-social-media-platforms-in-2020/}}, 52\% of American adults get news from digital platforms, out of which more than half of the people (53\%) said that they consume news from social media platforms. This is worrisome, especially during the time of COVID-19 pandemic\footnote{\url{https://www.pewresearch.org/fact-tank/2021/08/24/about-four-in-ten-americans-say-social-media-is-an-important-way-of-following-covid-19-vaccine-news/}} as most false claims regarding government rules, virus cures, vaccines, and more originate on various social media platforms, making users vulnerable to believing misinformation. Although these social media platforms have made efforts\footnote{\url{https://www.washingtonpost.com/technology/2020/11/09/facebook-twitter-election-misinformation-labels/}} to mitigate the spread of false narratives, it remains prevalent, as shown in this study and supported by previous research \citep{burel2020co}. 

Furthermore, people use different modalities of content such as text, images, videos, etc., to spread factually inaccurate claims. Figure \ref{fig:spatioFig4} (right) displays the transition in the modality of claims fact-checked in both query claim debunk and duplicate claim debunk. While the modality for text, video, and image remains consistent, there are also considerable cases where there is a transition between the modalities of content that state the same things.

\begin{figure}[!htbp]
    \centering
    \includegraphics[width=8cm,height=4.5cm]{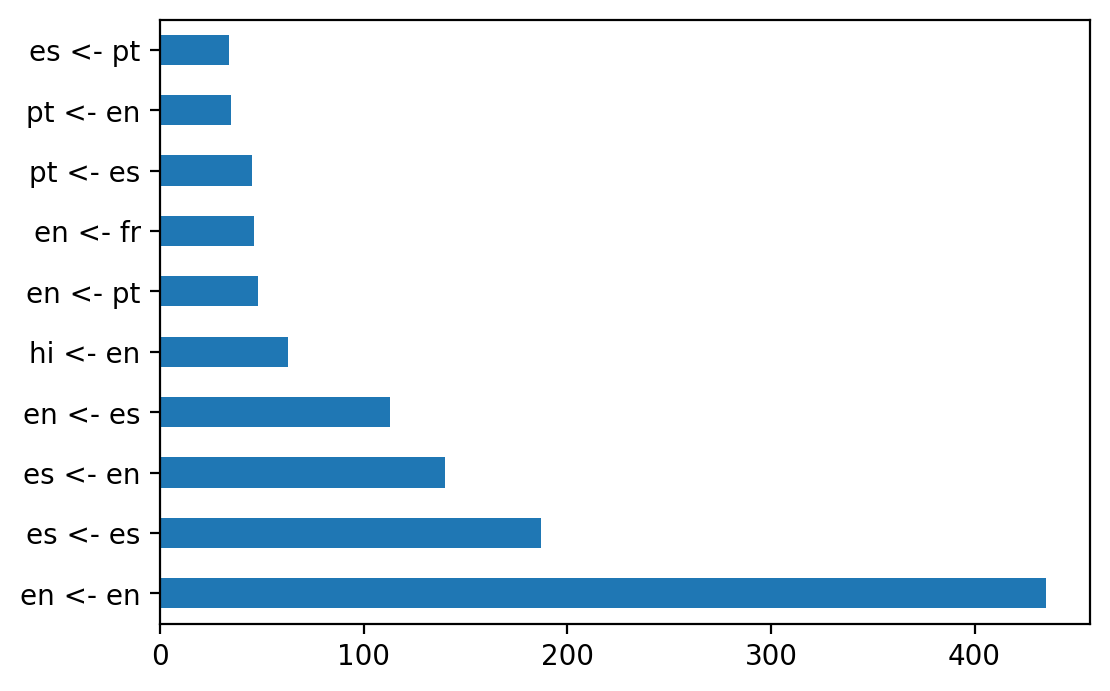}
    \caption{Top 10 count of cases showing the difference in the language used in the fact-checking articles for both the query claim debunk and the duplicate claim debunk. ISO-39 language code is used to denote the language.}
    \label{fig:spatioFig5}
\end{figure}

Figure \ref{fig:spatioFig5} shows the difference in the language used in the fact-checking articles for both the query claim debunk and the duplicate claim debunk for the top 10 language pairs. Here, the first symbol represents the ISO-39 language code of the query claim debunk, and the second one is the language used in duplicate claim debunk articles. It's noteworthy that for monolingual pairs, it's unusual to observe a significant number of duplicate claim debunks for which debunks already exist in the same language. Additionally, there are a considerable number of bilingual pairs, indicating the necessity for cross-lingual search before debunking a new claim, as discussed later in Finding 4.

\subsubsection*{Finding 3. COVID-19 misinformation involving general medical advice got spread across multiple countries and hence has the highest proportion of duplicate claim debunks in our dataset.}

To assist fact-checkers in quick debunking, prior work \citep{brennen2020types} categorised COVID-19 misinformation into various types, such as medical advice, virus origin, etc. We label the claims using CANTM model \citep{song2021classification} to understand which kinds of claims spread the most and have the highest number of duplicate claim debunks.

\begin{figure}[H]
    \centering
    \includegraphics[width=13cm,height=15cm]{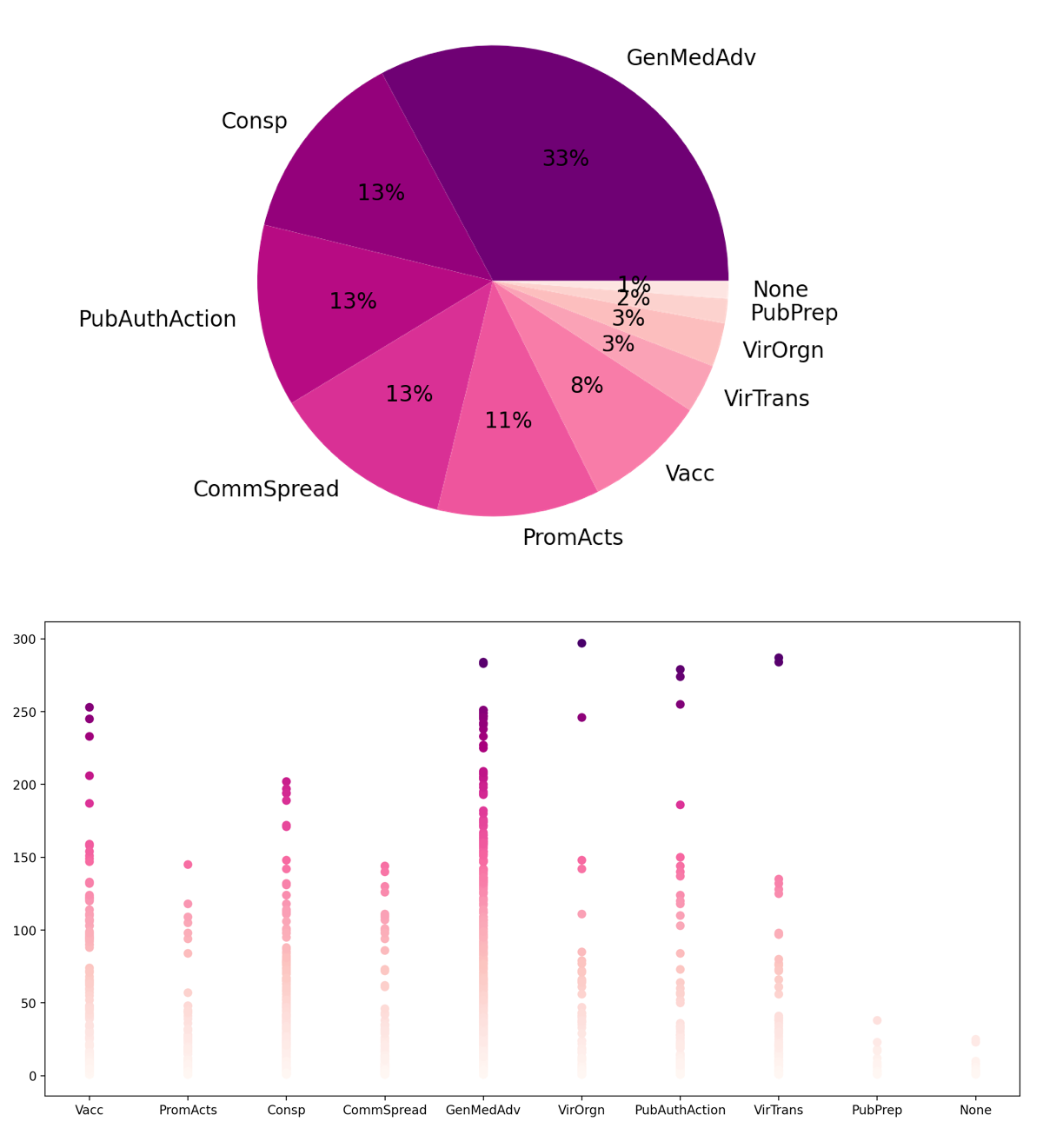}
    \caption{Top: Pieplot for categories of claims. Bottom: Difference in days between query claim debunks and duplicate claim debunks for different categories of claims. }
    \label{fig:spatioFig6}
\end{figure}

Figure \ref{fig:spatioFig6} (top) depicts a pie plot of the categories of claims for which multiple debunks exist. The COVID-19 misinformation categories include PubAuthAction (public authority), CommSpread (community spread and impact), GenMedAdv (medical advice, self-treatments, and virus effects), PromActs (prominent actors), Consp (conspiracies), VirTrans (virus transmission), VirOrgn (virus origin and properties), PubPrep (public reaction), Vacc (vaccines, medical treatments, and tests), and None (other). Misleading medical advice appears to be the most consistent topic of misinformation, accounting for the highest proportion at 33\%, followed by conspiracy theories, public authority actions, and community spread-based false claims, each making up 13\% of all cases. Overall, these recurring topics underscore the necessity for more efficient resource allocation to mitigate redundant debunking efforts.

Furthermore, Figure \ref{fig:spatioFig6} (bottom) is a scatter plot demonstrating the difference in days between query claim and duplicate claim debunks for different categories of claims. We observe that claims on general medical advice are most densely spread, indicating many cases where the publication date of duplicate claim debunks differs by several days. Claims about vaccines and conspiracy theories also exhibit a dense spread compared to others, which are denser on the lower end, depicting that the difference in days between the publication date of query claim and duplicate claim debunk is not much. 

In Table \ref{tab:spatioClassWords1}, we examine the top six words (after removing all non-useful words) in various categories of claims that have multiple debunks. Words such as ``Water'', ``lemon'' etc are most dominant in misinforming medical advice, while ``Honjo Tasuku'' and ``Gates'' can be observed in repeated claims involving conspiracy.

\begin{table}[!htbp]
\tiny
\centering

\begin{tabular}{lcccccccc}
\toprule
\textbf{Class} & \multicolumn{8}{c}{\textbf{Words}} \\ \midrule 
GenMedAdv & \cellcolor[HTML]{6FA8DC}water & \cellcolor[HTML]{E3EEF8}salt & \cellcolor[HTML]{E5F0F9}lemon & \cellcolor[HTML]{F0F6FC}cures & \cellcolor[HTML]{F0F6FC}breath & \cellcolor[HTML]{F8FBFE}vitamin & \cellcolor[HTML]{FFFFFF}vinegar & \cellcolor[HTML]{FFFFFF}tea \\
Consp & \cellcolor[HTML]{6FA8DC}nobel & \cellcolor[HTML]{99C1E6}honjo & \cellcolor[HTML]{ADCEEB}tasuku & \cellcolor[HTML]{C2DAF0}lab & \cellcolor[HTML]{C2DAF0}wuhan & \cellcolor[HTML]{D6E7F5}gates & \cellcolor[HTML]{FFFFFF}outbreak & \cellcolor[HTML]{FFFFFF}china \\
PubAuthAction & \cellcolor[HTML]{6FA8DC}people & \cellcolor[HTML]{93BEE5}china & \cellcolor[HTML]{A5C9EA}patients & \cellcolor[HTML]{A5C9EA}government & \cellcolor[HTML]{B7D4EE}police & \cellcolor[HTML]{C9DFF2}india & \cellcolor[HTML]{EDF5FB}court & \cellcolor[HTML]{FFFFFF}video \\
CommSpread & \cellcolor[HTML]{6FA8DC}people & \cellcolor[HTML]{D1E4F4}photo & \cellcolor[HTML]{D1E4F4}italy & \cellcolor[HTML]{E8F2FA}video & \cellcolor[HTML]{F4F9FD}patients & \cellcolor[HTML]{F4F9FD}china & \cellcolor[HTML]{FFFFFF}coffins & \cellcolor[HTML]{FFFFFF}victims \\
PromActs & \cellcolor[HTML]{6FA8DC}president & \cellcolor[HTML]{B7D4EE}ronaldo & \cellcolor[HTML]{C9DFF2}cristiano & \cellcolor[HTML]{EDF5FB}minister & \cellcolor[HTML]{EDF5FB}hospitals & \cellcolor[HTML]{FFFFFF}bill & \cellcolor[HTML]{FFFFFF}charles & \cellcolor[HTML]{FFFFFF}hotels \\
Vacc & \cellcolor[HTML]{6FA8DC}vaccine & \cellcolor[HTML]{F2F7FC}people & \cellcolor[HTML]{F6FAFD}cure & \cellcolor[HTML]{FBFDFE}bill & \cellcolor[HTML]{FBFDFE}gates & \cellcolor[HTML]{FDFEFF}dna & \cellcolor[HTML]{FDFEFF}russia & \cellcolor[HTML]{FFFFFF}pfizer \\
VirTrans & \cellcolor[HTML]{6FA8DC}hypoxia & \cellcolor[HTML]{6FA8DC}masks & \cellcolor[HTML]{9FC5E8}use & \cellcolor[HTML]{CFE2F4}mask & \cellcolor[HTML]{E7F1FA}chicken & \cellcolor[HTML]{FFFFFF}flu & \cellcolor[HTML]{FFFFFF}creator & \cellcolor[HTML]{FFFFFF}pcr \\
VirOrgn & \cellcolor[HTML]{6FA8DC}video & \cellcolor[HTML]{9FC5E8}wuhan & \cellcolor[HTML]{CFE2F4}virus & \cellcolor[HTML]{CFE2F4}china & \cellcolor[HTML]{CFE2F4}market & \cellcolor[HTML]{E7F1FA}bats & \cellcolor[HTML]{FFFFFF}chicken & \cellcolor[HTML]{FFFFFF}hubei \\
PubPrep & \cellcolor[HTML]{6FA8DC}people & \cellcolor[HTML]{B7D4EE}lions & \cellcolor[HTML]{DBEAF7}streets & \cellcolor[HTML]{DBEAF7}russia & \cellcolor[HTML]{EDF5FB}homes & \cellcolor[HTML]{EDF5FB}masks & \cellcolor[HTML]{FFFFFF}berlin & \cellcolor[HTML]{FFFFFF}pandemic \\ \bottomrule
\end{tabular}

\caption{Top six words in different categories of claims that have multiple debunks; darker blue means higher volume.}
\label{tab:spatioClassWords1}
\end{table}

\begin{figure}[H]
    \centering
    \includegraphics[width=14.2cm,height=4cm]{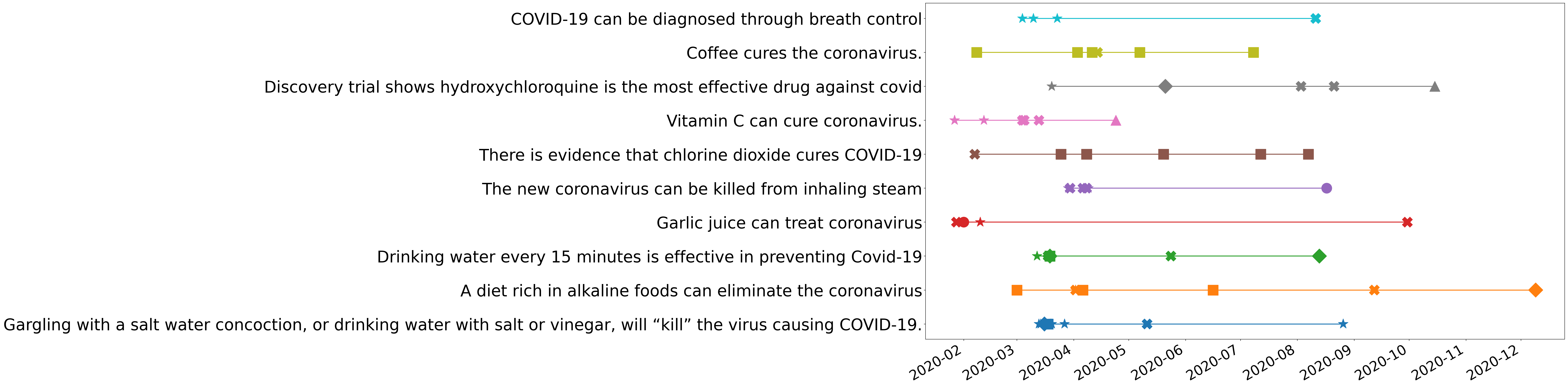}
    \caption{Timeline for a sample of 10 claims about fallacious medical advice. Here the language of debunk article is denoted by different symbols like English: $\bigstar$; Spanish: $\blacksquare$; Hindi: $\bullet$; Portuguese: $\blacklozenge$; French: $\blacktriangle$; Other: $\textbf{X}$; }
    \label{fig:spatioFig7}
\end{figure}

We further examine claims that are widely spread and have debunks published at different times of the year. Figure \ref{fig:spatioFig7} presents a sample of 10 false claims about fallacious medical advice, including cures, remedies, and prevention methods specific to COVID-19. We find that the duplicate claim debunks for these claims are spread across the entire year and are published in different languages.

\begin{figure}[H]
    \centering
    \includegraphics[width=14.2cm,height=12cm]{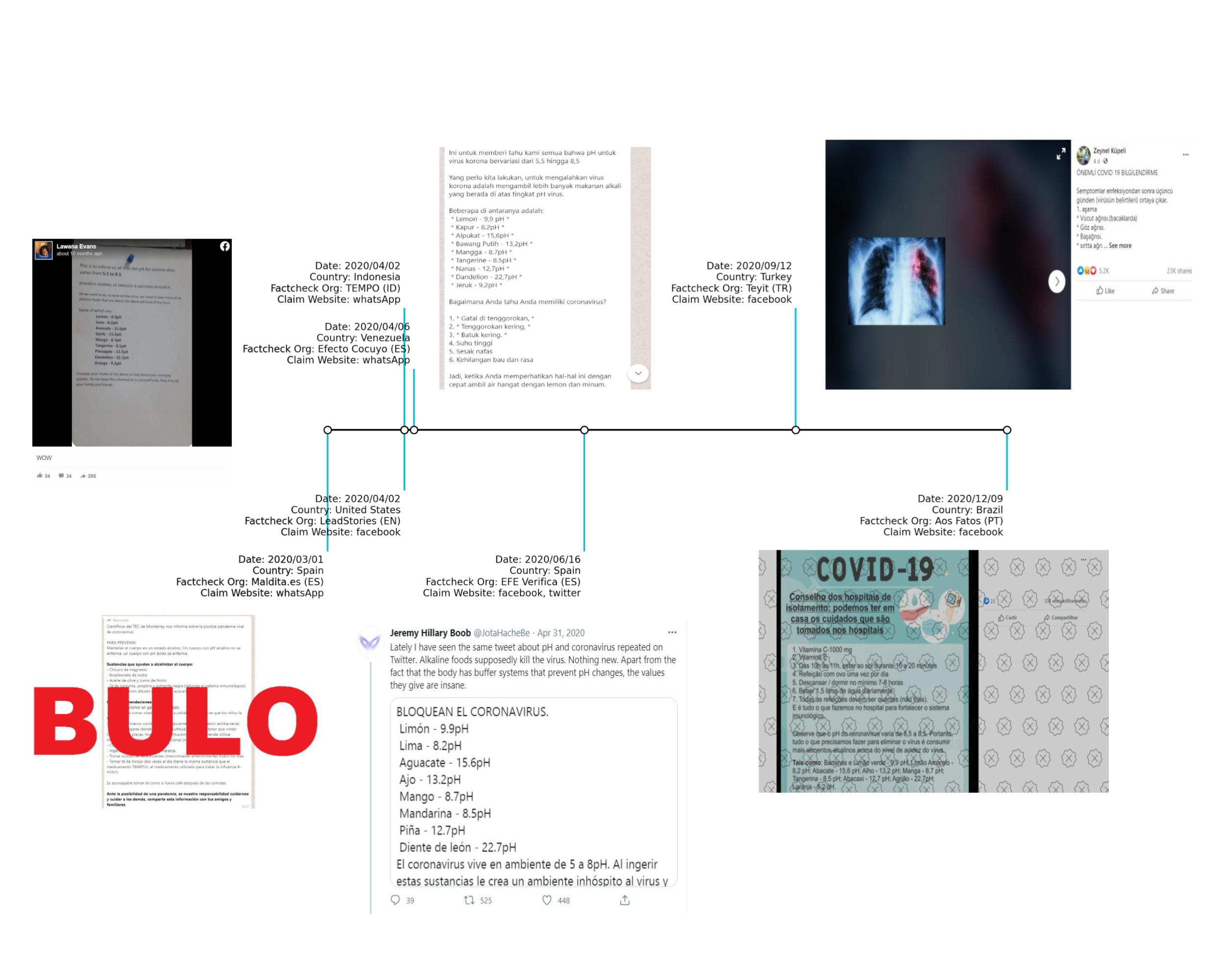}
    \caption{A detailed timeline of claim: “A diet rich in alkaline foods can eliminate the coronavirus”. All the images show the same claims being spread on different social media websites in different languages and varied modalities (top left and bottom right are the images shared on Facebook; top right and bottom centre are the images accompanied by some text shared on Facebook and Twitter respectively; top centre and bottom left show text shared on WhatsApp)}
    \label{fig:spatioFig8}
\end{figure}

Subsequently, Figure \ref{fig:spatioFig8} illustrates the timeline of debunks for claims about the consumption of an alkaline-rich diet to eliminate the coronavirus. From our dataset, it appears that the claim was first debunked in Spain in March 2020 and after a month a similar claim was debunked in Indonesia and the United States but it was still here to stay. It is surprising and yet worrisome that the same claim was again debunked in Turkey and Brazil in September and December respectively. One thing that might have led to this unknowing spread of previously debunked claims is the language of the fact-checking article, as they all differ (shown in Figure \ref{fig:spatioFig8} with ISO-39 language codes enclosed in brackets after the name of the fact-check organisation).

We also investigate the language and modality of the claims and find that claims written in one language are sometimes transformed into other languages and varied modalities (eg. text to image) before being propagated to other countries. The social media platforms used to spread the claim in different countries also change over time. Figure \ref{fig:spatioFig8} shows that the same claim was shared on Facebook, WhatsApp, and Twitter. \\

\begin{figure}[H]
    \centering
    \includegraphics[width=14.2cm,height=4cm]{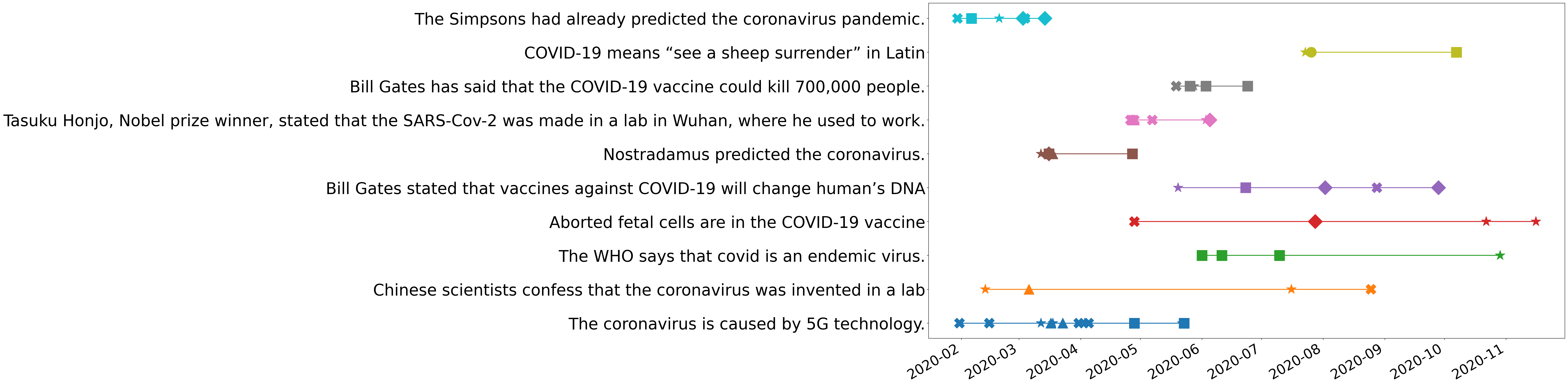}
    \caption{Timeline for a sample of 10 conspiracy theories concerning COVID-19.  Here the language of the debunk article is denoted by different symbols like English: $\bigstar$; Spanish: $\blacksquare$; Hindi: $\bullet$; Portuguese: $\blacklozenge$; French: $\blacktriangle$; Other: $\textbf{X}$ }
    \label{fig:spatioFig9}
\end{figure}

\begin{figure}[H]
    \centering
    \includegraphics[width=13.2cm,height=5cm]{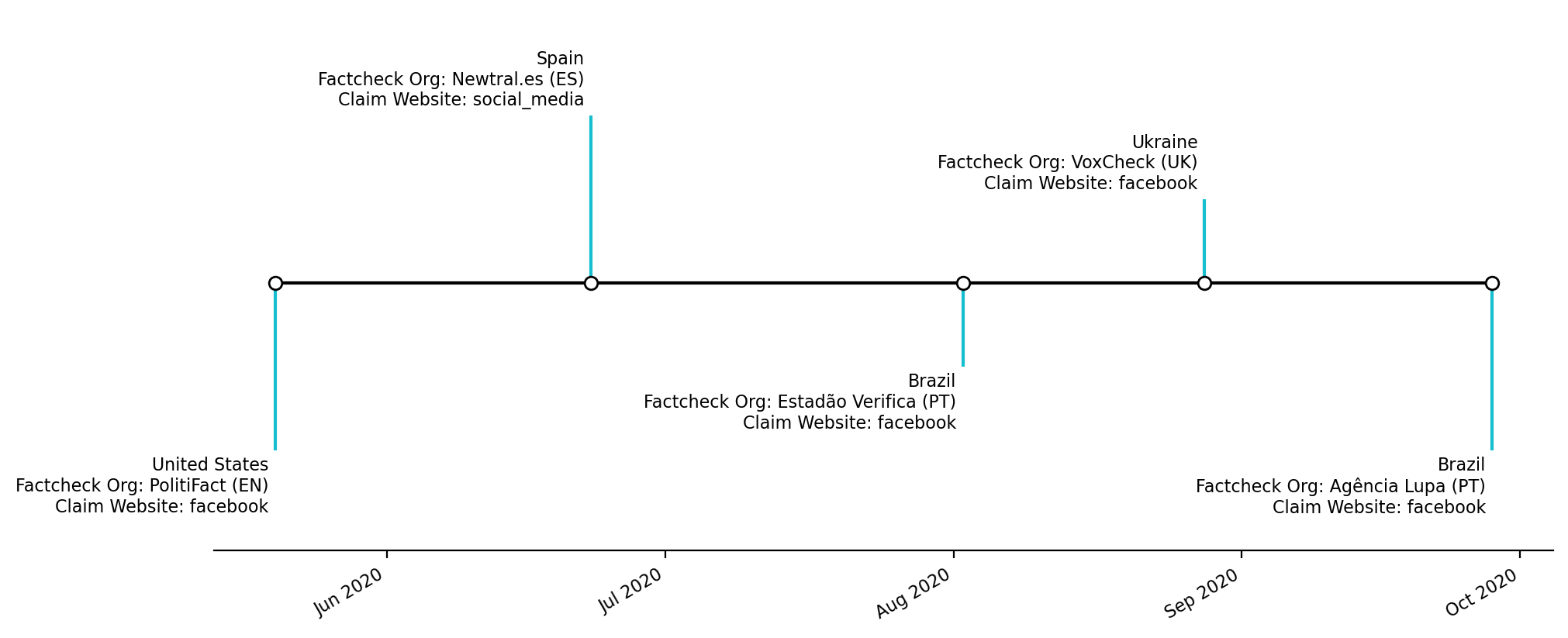}
    \caption{Detailed timeline of claim: ``Bill Gates stated that vaccines against COVID-19 will change human’s DNA''}
    \label{fig:spatioFig10}
\end{figure}

Figure \ref{fig:spatioFig9} illustrates conspiracy theories that have been debunked multiple times. The belief that COVID-19 is linked to 5G technology was common across many countries, despite having been debunked before. Additionally, there are numerous falsely attributed claims and conspiracies involving Bill Gates. For instance, Figure \ref{fig:spatioFig10} displays the timeline of debunks about claims alleging a statement from Bill Gates that the COVID-19 vaccine can change human DNA. All the debunks appear in multiple languages at different times over the time span of five months from June to October 2020.

\subsubsection*{Finding 4. The IFCN database mainly consists of cases where there isn't even a single duplicate claim debunk in the same language as that of the query claim debunk, highlighting the necessity of including multilingual debunk search in the fact-checking pipeline.}

As mentioned earlier, we identified 1,070 debunks about claims that already have debunks published by an IFCN fact-checker. Among these 1,070 cases, there are a total of 627 (59\%) instances for which we don't have a single duplicate claim debunk in the same language as that of the query claim debunk. Alternatively, this shows that if a person from some country is willing to search for fact-check articles about a claim that has already been debunked in a language different from what the person understands, then he/she might not be able to do so due to the language barrier. Although one can make efforts to search through the content in multiple languages, it’s usually not done because it’s inefficient and it’s probably the reason claims spread, incognizant of the fact that they have already been debunked in the past. Therefore, the need for multilingual and cross-lingual debunk search in the initial stages of the fact-checking pipeline becomes imperative.

Before delving into fact-checking a claim, it is crucial to check whether the claim or its equivalent has already been debunked by a fact-checking organisation in a different language. While there are commercially available debunk database search tools by Google\footnote{\url{https://toolbox.google.com/factcheck/explorer}} and WeVerify\footnote{\url{https://weverify.eu/}}, to the best of our knowledge, these tools are limited to monolingual search.
Our analysis highlights the need for a cross-lingual/multilingual retrieval search, where a comprehensive pool of debunked narratives from around the world is considered, irrespective of the language used in the fact-checking article. Given the time-consuming nature of manual fact-checking, avoiding duplicated efforts in debunking narratives that have already been debunked in the past is paramount. Therefore, the ability to search for previously debunked narratives in multiple languages is beneficial for fact-checkers.

On the other hand, while it may be impossible to fact-check every claim, social media platforms can take the initiative to warn users before they share content containing previously debunked narratives. Over the years, numerous fact-checking organisations have emerged, accumulating a vast corpus of fact-checking articles \citep{augenstein2019multifc, shahi2020fakecovid, gupta2021x} debunking various claims in different languages. This data can be effectively utilised to quickly debunk repeated false narratives appearing on various social media platforms, thereby limiting their spread and potential harm.

\section{Limitations and Future Work}

Our work should be seen in light of the following limitations: i) For all the fact-checking articles debunking similar narratives, we did not consider any changes in rulings made by fact-checkers over time. In other words, we assume that if a claim is initially declared false by some fact-checking organisation, then it remains false irrespective of the time or place of debunking of a similar claim. This is something we plan to investigate in detail in our future work. ii) While the dataset utilised in our analysis may be considered weakly labelled, we mitigate this limitation by leveraging state-of-the-art semantic similarity models with a high threshold. Additionally, we conduct manual checks to ensure that only relevant duplicate claim debunks are included in our study. iii) Finally, we presume that the spread of debunked narratives is due to the spreader being unaware of the previously debunked article about a similar narrative that got spread in the past. The main aim of this study is to draw attention to the general public and fact-checkers regarding the presence of duplicate claim debunks, suggesting ways to mitigate the spread of debunked narratives and better deal with potential infodemics in the future.

\section{Conclusion}
\label{spatioconclude}

The onset of the COVID-19 pandemic led to a global infodemic that has brought unprecedented challenges for citizens, media, and fact-checkers worldwide. To address this challenge, over a hundred fact-checking initiatives worldwide have been monitoring the information space in their countries and publishing regular debunks of viral false COVID-19 narratives. In this paper, we examine the database of the CoronaVirusFacts Alliance, which contains 10,381 debunks related to COVID-19 published in multiple languages by different fact-checking organisations. Our spatiotemporal analysis reveals that similar or nearly duplicate false COVID-19 narratives have been spreading in multiple modalities and on various social media platforms in different countries, sometimes as much as several months after the first debunk of that narrative has been published by an IFCN fact-checker. We also find that misinformation involving general medical advice has spread across multiple countries and hence has the highest proportion of false COVID-19 narratives that keep being debunked. Furthermore, as manual fact-checking is an onerous task in itself, therefore the need to repeatedly debunk the same narrative in different countries is leading, over time, to a significant waste of fact-checker resources. In response, we advocate for the incorporation of a multilingual debunked narrative search tool into the fact-checking pipeline. Additionally, we strongly recommend that social media platforms adopt this technology at scale, so as to make the best use of scarce fact-checker resources.

\section*{Funding} 
This work was partially funded by the WeVerify and SoBigData++ projects (EUH2020, grant agreements:
825297 and 871042).

\section*{Data Availability} 
The International Fact-Checking Network (IFCN) debunks used in this paper are publicly available at
https://www.poynter.org/ifcn-covid-19-misinformation/ and the code to scrape the debunks is available at
https://github.com/iknoorjobs/IFCN-scraper

\bibliographystyle{apalike}
\vspace*{-0.2cm}
\bibliography{references}

\begin{thebibliography}{}

\bibitem[Augenstein et~al., 2019]{augenstein2019multifc}
Augenstein, I., Lioma, C., Wang, D., Chaves~Lima, L., Hansen, C., Hansen, C., and Simonsen, J.~G. (2019).
\newblock {M}ulti{FC}: A real-world multi-domain dataset for evidence-based fact checking of claims.
\newblock In {\em Proceedings of the 2019 Conference on Empirical Methods in Natural Language Processing and the 9th International Joint Conference on Natural Language Processing (EMNLP-IJCNLP)}, pages 4685--4697, Hong Kong, China. Association for Computational Linguistics.

\bibitem[Barr{\'o}n-Cedeno et~al., 2020]{barron2020overview}
Barr{\'o}n-Cedeno, A., Elsayed, T., Nakov, P., Da~San~Martino, G., Hasanain, M., Suwaileh, R., Haouari, F., Babulkov, N., Hamdan, B., Nikolov, A., et~al. (2020).
\newblock Overview of checkthat! 2020: Automatic identification and verification of claims in social media.
\newblock In {\em International Conference of the Cross-Language Evaluation Forum for European Languages}, pages 215--236. Springer.

\bibitem[Bontcheva et~al., 2020]{Bontcheva2020}
Bontcheva, K., Posetti, J., Teyssou, D., Meyer, T., Gregory, S., Hanot, C., and Maynard, D. (2020).
\newblock Balancing act: Countering digital disinformation while respecting freedom of expression.
\newblock Technical report, United Nation Educational, Scientific and Cultural Organization.

\bibitem[Brennen et~al., 2020]{brennen2020types}
Brennen, J.~S., Simon, F., Howard, P.~N., and Nielsen, R.~K. (2020).
\newblock Types, sources, and claims of covid-19 misinformation.
\newblock {\em Reuters Institute}, 7(3):1.

\bibitem[Bridgman et~al., 2020]{bridgman2020causes}
Bridgman, A., Merkley, E., Loewen, P.~J., Owen, T., Ruths, D., Teichmann, L., and Zhilin, O. (2020).
\newblock The causes and consequences of covid-19 misperceptions: Understanding the role of news and social media.
\newblock {\em Harvard Kennedy School Misinformation Review}, 1(3).

\bibitem[Burel et~al., 2020]{burel2020co}
Burel, G., Farrell, T., Mensio, M., Khare, P., and Alani, H. (2020).
\newblock Co-spread of misinformation and fact-checking content during the covid-19 pandemic.
\newblock In {\em International Conference on Social Informatics}, pages 28--42. Springer.

\bibitem[Ecker et~al., 2010]{ecker2010explicit}
Ecker, U.~K., Lewandowsky, S., and Tang, D.~T. (2010).
\newblock Explicit warnings reduce but do not eliminate the continued influence of misinformation.
\newblock {\em Memory \& cognition}, 38:1087--1100.

\bibitem[FullFact, 2024]{fullfactarticle}
FullFact (2024).
\newblock Fullfact report tracks fake covid-19 news across five countries – society of editors.

\bibitem[Gupta and Srikumar, 2021]{gupta2021x}
Gupta, A. and Srikumar, V. (2021).
\newblock {X}-fact: A new benchmark dataset for multilingual fact checking.
\newblock In {\em Proceedings of the 59th Annual Meeting of the Association for Computational Linguistics and the 11th International Joint Conference on Natural Language Processing (Volume 2: Short Papers)}, pages 675--682, Online. Association for Computational Linguistics.

\bibitem[Lewandowsky et~al., 2012]{lewandowsky2012misinformation}
Lewandowsky, S., Ecker, U.~K., Seifert, C.~M., Schwarz, N., and Cook, J. (2012).
\newblock Misinformation and its correction: Continued influence and successful debiasing.
\newblock {\em Psychological science in the public interest}, 13(3):106--131.

\bibitem[Limaye et~al., 2020]{limaye2020building}
Limaye, R.~J., Sauer, M., Ali, J., Bernstein, J., Wahl, B., Barnhill, A., and Labrique, A. (2020).
\newblock Building trust while influencing online covid-19 content in the social media world.
\newblock {\em The Lancet Digital Health}, 2(6):e277--e278.

\bibitem[Liu et~al., 2019]{liu2019roberta}
Liu, Y., Ott, M., Goyal, N., Du, J., Joshi, M., Chen, D., Levy, O., Lewis, M., Zettlemoyer, L., and Stoyanov, V. (2019).
\newblock {RoBERTa}: A robustly optimized bert pretraining approach.
\newblock {\em ArXiv preprint}, abs/1907.11692.

\bibitem[McGlynn et~al., 2020]{mcglynn2020misinformation}
McGlynn, J., Baryshevtsev, M., and Dayton, Z.~A. (2020).
\newblock Misinformation more likely to use non-specific authority references: Twitter analysis of two covid-19 myths.
\newblock {\em Harvard Kennedy School Misinformation Review}, 1(3).

\bibitem[Nakov, 2020]{nakov2020can}
Nakov, P. (2020).
\newblock Can we spot the" fake news" before it was even written?
\newblock {\em ArXiv preprint}, abs/2008.04374.

\bibitem[Nogueira and Cho, 2019]{nogueira2019passage}
Nogueira, R. and Cho, K. (2019).
\newblock Passage re-ranking with bert.
\newblock {\em ArXiv preprint}, abs/1901.04085.

\bibitem[Nyhan and Reifler, 2010]{nyhan2010corrections}
Nyhan, B. and Reifler, J. (2010).
\newblock When corrections fail: The persistence of political misperceptions.
\newblock {\em Political Behavior}, 32(2):303--330.

\bibitem[Pillai and Fazio, 2021]{pillai2021effects}
Pillai, R.~M. and Fazio, L.~K. (2021).
\newblock The effects of repeating false and misleading information on belief.
\newblock {\em Wiley Interdisciplinary Reviews: Cognitive Science}, page e1573.

\bibitem[Posetti and Bontcheva, 2020]{Posetti2020}
Posetti, J. and Bontcheva, K. (2020).
\newblock Policy brief 1, disinfodemic: Deciphering covid-19 disinformation.
\newblock Technical report, United Nation Educational, Scientific and Cultural Organization.

\bibitem[Reis et~al., 2020]{reis2020can}
Reis, J., Melo, P. d.~F., Garimella, K., and Benevenuto, F. (2020).
\newblock Can whatsapp benefit from debunked fact-checked stories to reduce misinformation?
\newblock {\em ArXiv preprint}, abs/2006.02471.

\bibitem[Shaar et~al., 2020]{shaar-etal-2020-known}
Shaar, S., Babulkov, N., Da~San~Martino, G., and Nakov, P. (2020).
\newblock That is a known lie: Detecting previously fact-checked claims.
\newblock In {\em Proceedings of the 58th Annual Meeting of the Association for Computational Linguistics}, pages 3607--3618, Online. Association for Computational Linguistics.

\bibitem[Shahi and Nandini, 2020]{shahi2020fakecovid}
Shahi, G.~K. and Nandini, D. (2020).
\newblock Fakecovid--a multilingual cross-domain fact check news dataset for covid-19.
\newblock {\em ArXiv preprint}, abs/2006.11343.

\bibitem[Singh et~al., 2020]{singh2020coherence}
Singh, I., Deepak, P., and Anoop, K. (2020).
\newblock On the coherence of fake news articles.
\newblock In {\em Joint European Conference on Machine Learning and Knowledge Discovery in Databases}, pages 591--607. Springer.

\bibitem[Singh et~al., 2021]{singh2021multistage}
Singh, I., Scarton, C., and Bontcheva, K. (2021).
\newblock Multistage bicross encoder for multilingual access to covid-19 health information.
\newblock {\em PloS one}, 16(9):e0256874.

\bibitem[Song et~al., 2021]{song2021classification}
Song, X., Petrak, J., Jiang, Y., Singh, I., Maynard, D., and Bontcheva, K. (2021).
\newblock Classification aware neural topic model for covid-19 disinformation categorisation.
\newblock {\em PloS one}, 16(2):e0247086.

\bibitem[Tasnim et~al., 2020]{tasnim2020impact}
Tasnim, S., Hossain, M.~M., and Mazumder, H. (2020).
\newblock Impact of rumors and misinformation on covid-19 in social media.
\newblock {\em Journal of preventive medicine and public health}, 53(3):171--174.

\bibitem[Thorne and Vlachos, 2018]{thorne2018automated}
Thorne, J. and Vlachos, A. (2018).
\newblock Automated fact checking: Task formulations, methods and future directions.
\newblock {\em arXiv preprint arXiv:1806.07687}.

\bibitem[Zhou and Zafarani, 2020]{zhou2020survey}
Zhou, X. and Zafarani, R. (2020).
\newblock A survey of fake news: Fundamental theories, detection methods, and opportunities.
\newblock {\em ACM Computing Surveys (CSUR)}, 53(5):1--40.

\end{thebibliography}

\section{Appendix}

\subsection{Gephi Plot}
\label{gephi}
\begin{figure}[!htbp]
    \centering
    \includegraphics[width=15cm,height=13cm]{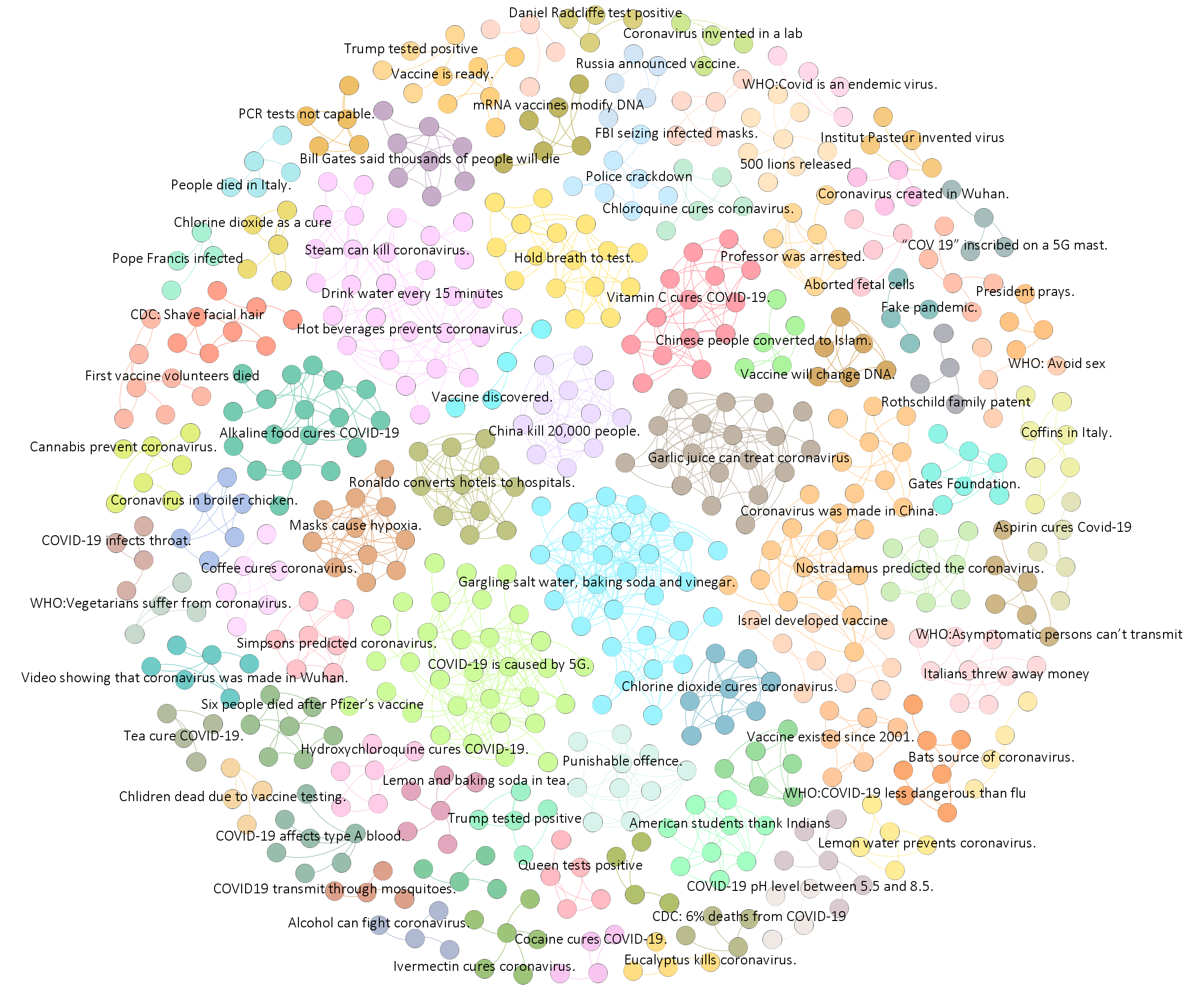}
    \caption{Cluster visualisation for duplicate claim debunks.}
    \label{fig:gephi}
\end{figure}

Out of 10,381 debunks in the IFCN database, we find 1070 debunked claims that already had a debunk about the same false narrative from a different fact-checking organisation in the past. We clustered together all such duplicate claim debunks which have more than three debunks that fact-check similar claims and produced a GRAPHML-file to visualise the clusters using java-based network analysis applications such as Gephi (Figure \ref{fig:gephi}). The Fructhterman-Reingold force-directed graph drawing algorithm is used to visualise the network in a compact circle with coloured cluster separation based on the modularity class. Here, a node represents a debunk from the fact-checking organisation and the colour represents the cluster of all duplicate claim debunks. The claim statement for each cluster is mentioned as shown in Figure \ref{fig:gephi}.

\end{document}